\documentclass[aps,prd,tighten,showpacs]{revtex4}

\usepackage{graphicx}
\usepackage{bm}
\usepackage{amsmath,amssymb}
\usepackage{latexsym}

\newcommand{\pp}{\phantom}

\begin{document}

\title{Vacuum polarization of the quantized massive scalar field in Reissner-Nordstr\"om  spacetime}

\author{Jerzy Matyjasek, Dariusz Tryniecki and Katarzyna Zwierzchowska}
\affiliation{Institute of Physics, Maria Curie-Sk\l odowska University\\
pl. Marii Curie-Sk\l odowskiej 1, 20-031 Lublin, Poland}

\date{\today}

\begin{abstract}
The approximation of the renormalized stress-energy tensor of the quantized
massive scalar field in Reissner-Nordstr\"om spacetime is constructed. It
is achieved by functional differentiation of the first two nonvanishing
terms of the Schwinger-DeWitt expansion involving the coincidence limit of
the Hadamard-Minakshisundaram-DeWitt-Seely coefficients $[a_{3}]$ and
$[a_{4}]$ with respect to the metric tensor. It is shown, by comparison
with the existing numerical results, that inclusion of the second-order
term leads to substantial improvement of the approximation of the exact
stress-energy tensor. The approximation to the field fluctuation, 
$\langle \phi^{2} \rangle,$ is constructed and briefly discussed 

\end{abstract}





\maketitle
\section{\label{intro}Introduction}

Recently, there has been a renewal of interest in calculations of the
coincidence limits of the covariant derivatives of various bitensors and
bispinors, such as the world function $\sigma (x,x'),$  van Vleck-
Morette determinant, bivectors and bispinors of the parallel displacement
and the objects constructed form them, with the special emphasis put on the
diagonal Hadamard-Minakshisundaram-DeWitt-Seely (HMDS) coefficients $[a_{n}(x,x')]$ and 
the renormalized stress-energy tensor~\cite{kocio1,kocio2,lemos,kocio3,ottewill}. 
This has largely been stimulated by new methods of computations as well as 
improvement of the computer algebra and the recent findings substantially extended 
previous results~\cite{Bryce1,sakai,gi11,gi12,Parker,AvraPhD,Avraros,Avranuc,Amsterdamski,Ven}.

In this article we construct the approximation to the renormalized stress-
energy tensor of the massive scalar field with arbitrary curvature coupling
in the spacetime of the Reissner-Nordstr\"om black hole. This tensor has
been the subject of previous studies both in the massless and massive
cases. The extensive numerical calculations have been reported in Ref.~\cite{AHS95},
where, additionally, the first-order approximation in the large mass limit
(which consists of the terms proportional to $m^{-2}$) has been developed.
These calculations were based on the sixth-order WKB approximation of the
solutions of the scalar field equations and the summation of the thus
obtained mode functions. The latter result has been reconstructed with the
aid of the Schwinger-DeWitt approximation of the renormalized effective
action, $W_{R},$  and subsequently  generalized to the spinor and vector
fields in  the $R = 0$ geometries in Ref.~\cite{kocio1}. The approximate stress-
energy tensor in a general background geometry has been constructed in
Ref.~\cite{kocio2}. A careful analysis carried out
in~Ref.\cite{semiRN} demonstrated that the approximation is reasonable for
$M m>2$ and should become increasingly accurate as the ratio $\lambda_{C}
/{\cal L}$ decreases, leaving, however,  room for further improvement. Here
$M$ is the mass of the Reissner-Nordstr\"om black hole, $\lambda_{C}$ is
the Compton length associated with the massive field and ${\cal L}$ is the
characteristic  length of the background geometry.

Our aim is to provide a better approximation than those proposed in 
Refs.~\cite{AHS95,kocio1,kocio2} and this is achieved by the inclusion 
of all relevant terms of the background dimensionality 8 (which equals
twice the order of the coefficient $[a_{n}]$ or the total number of derivatives of the
metric tensor in each term) in the renormalized effective action. Such
terms are proportional to $m^{-4}$ and constitute the next-to-leading order
of the approximation. The approximate stress-energy tensor can be obtained
by the  functional differentiation of the thus constructed effective action
with respect to the metric tensor. 

The basic building blocks of the renormalized one-loop $W_{R}$ are the coincidence 
limits of the HMDS coefficients
which are local quantities constructed from the Riemann tensor, its
covariant derivatives and contractions. The spin of the field and the type
of the curvature coupling enters $W_{R}$ through the numerical coefficients
and  the background dimensionality of $[a_{3}]$ and $[a_{4}]$ is 6 and 8,
respectively. In general, the coefficient $[a_{n}]$ is a linear combination 
of the Riemann monomials and belongs to $\bigoplus_{q=1}^{n} {\cal R}_{2n,q}^{0},$
where ${\cal R}_{s,q}^{r}$ is a vector space of Riemannian polynomials of
rank $r$ (the number of free tensor indices), degree $q$ (number of factors)
and order $s$ (number of derivatives).

The calculation of the functional derivatives of the effective action with
respect to $g_{ab}$ is rather tedious and time consuming process.
Fortunately, for the spherically- symmetric geometries  one can
considerably simplify calculations.  Indeed, substituting the line element
of the general static and spherically-symmetric spacetime expressed in the
Schwarzschild gauge into the effective action and performing simple
integrations one obtains a reduced functional that depends on the two
metric potentials. Two components of the stress-energy tensor are given by
appropriate Lagrange derivatives of the Lagrangian of the reduced action
functional with respect to the time and radial components of the metric
tensor. This approach is justified by the symmetric criticality theorems of
Palais~\cite{Palais,Deser1} and the remaining components can easily be
calculated form the covariant conservation equation.

The coefficients $[a_{n}]$ are also the basic building blocks of the approximate
field fluctuation $\langle \phi^{2}\rangle$ and the knowledge of $[a_{2}]$,
$[a_{3}]$ and $[a_{4}]$ allows for detailed analysis of the role played by
the next-to-leading and the next-to-next-to-leading terms. One expects
that some general features exhibited by $\langle \phi^{2}\rangle$ 
are also shared by the stress-energy tensor and if so it would be a fortunate 
circumstance.

The paper is organized as follows. In Sec. II the basic building blocks of
the approximation, i.e., the coincidence limits of the HMDS
coefficients $[a_{3}]$ and $[a_{4}]$  as calculated within the framework of
the covariant DeWitt method are presented in maximally condensed form. The
renormalized expectation value $\langle \phi^{2}\rangle$ in the Reissner-
Nordstr\"om geometry is computed in Sec. III. The effective action and the
stress-energy tensor of the quantized massive scalar fields in the
Reissner-Nordstr\"om geometry is constructed and discussed in Sec. IV. A
comparison with the numeric results indicate that inclusion of the next-to-
leading term substantially improves approximation.

\section{Hadamard-Minakshisundaram-DeWitt-Seely coefficients $[a_{3}]$ and $[a_{4}].$}

It is a well-known fact that for sufficiently massive quantized fields, i.e., when
the Compton length $\lambda_{C}$ is smaller than the characteristic radius
of curvature of the background geometry, the asymptotic expansion of the effective action in powers of
$m^{-2}$ may be used to describe various physical phenomena. It is because
the nonlocal contribution to the effective action can be neglected whereas
the vacuum polarization part is determined by the local geometry. The
renormalized effective action constructed within the framework of the
Schwinger-DeWitt approximation for the quantized massive scalar field
satisfying the covariant Klein-Gordon equation with the curvature coupling
$\xi$
\begin{equation}
     \left( \Box \,-\,\xi R\,-\,m^{2}\right) \phi \,=\,0,  
                                                  \label{wave}
\end{equation}
can be written in the form
\begin{equation}
W_{R}\,=\,\frac{1}{32\pi^{2}}\sum_{n=3}^{\infty}
\frac{(n-3)!}{(m^{2})^{n-2}}\int d^{4}x \sqrt{g} [a_{n}],
                           \label{Weff}
\end{equation}
where $[a_{n}]$ is constructed from  the Riemann tensor, its covariant derivatives 
up to $2n-2$ order and appropriate contractions. For the technical details of this 
approach the reader is referred, for example, to Refs.~\cite{Barvinsky:1985an,FZ3} 
and the references cited therein.

Inspection of Eq.~(\ref{Weff}) shows that the lowest term of the
approximate $W_{R}$ is to be constructed from the (integrated) coincidence
limit of the fourth HMDS coefficient,
$[a_{3}],$ whereas the next to leading term is constructed form $[a_{4}].$
Here we will confine ourselves to the first two  terms of the
expansion~(\ref{Weff}). 

The coefficients $a_{n}(x,x')$ satisfy the equation
\begin{equation}
\sigma^{;i} a_{n;i} + n a_{n} - \Delta^{-1/2}\Box \left( \Delta^{1/2} 
a_{n-1}\right) + \xi R a_{n-1} =0,
                                                      \label{DeWitt}
\end{equation}
with the boundary condition $a_{0}(x,x')=1.$ Here $\Delta(x,x')$ is  
the van Vleck-Morette determinant and the biscalar $\sigma(x,x')$ is 
equal to one half the square of the distance along the geodesic
between $x$ and $x'.$
From the recursive relation (\ref{DeWitt}) it is clear that to construct,
say, $[a_{4}],$ one has to calculate coincidence limits of
$a_{3},$ $a_{3;i_{1}}$ and $a_{3; i_{1} i_{2}},$
which, in turn, require calculation of $[a_{2}]$ to $[a_{2; i_{1}...i{4}}],$
and so forth.
Both $[a_{3}]$ and $[a_{4}]$  used in this paper
have been calculated within the framework of the manifestly covariant method
proposed by DeWitt with the aid of the FORM and its multithread version
TFORM~\cite{Vermaseren,Vermaseren-Tentyukov}. 
Further simplifications,
after appropriate syntax conversion, were carried out with the aid of the
package INVAR~\cite{JoseGM+P,JoseGM+P+M}.

Since the coefficients $[a_{3}]$ and
$[a_{4}]$ are the basis of the present calculations we shall display them
at length.
The coefficient $[a_{3}],$ when simplified with the aid of the INVAR, can
be written in the form
\begin{equation}
[a_{3}] = a_{3}^{(0)} + \xi a_{3}^{(1)} + \xi^{2} a_{3}^{(2)} + \xi^3 a_{3}^{(3)},
                                                                 \label{a3a}
\end{equation}
where
\begin{eqnarray}
a_{3}^{(0)} &=& \frac{11}{1680}  R^3+
\frac{17}{5040}  R_{;a}^{\pp{;\pp{a}}}  R_{\pp{;\pp{a}}}^{;a}-
\frac{1}{2520}  R_{ab;c}^{\pp{a}\pp{b}\pp{;\pp{c}}}  R_{\pp{a}\pp{b}\pp{;\pp{c}}}^{ab;c}-
\frac{1}{1260}  R_{ab;c}^{\pp{a}\pp{b}\pp{;\pp{c}}}  R_{\pp{a}\pp{c}\pp{;\pp{b}}}^{ac;b}\nonumber \\ 
 &+
&\frac{1}{560}  R_{abcd;e}^{\pp{a}\pp{b}\pp{c}\pp{d}\pp{;\pp{e}}}  
R_{\pp{a}\pp{b}\pp{c}\pp{d}\pp{;\pp{e}}}^{abcd;e}+
\frac{1}{180}  R  R_{;a\pp{a}}^{\pp{;\pp{a}}a}+
\frac{1}{280}  R_{;a\pp{a}b\pp{b}}^{\pp{;\pp{a}}a\pp{b}b}+
\frac{1}{420}  R_{;ab}^{\pp{;\pp{a}}\pp{b}}  R_{\pp{a}\pp{b}}^{ab}\nonumber \\ 
 &-
&\frac{1}{630}  R_{ab;c\pp{c}}^{\pp{a}\pp{b}\pp{;\pp{c}}c}  R_{\pp{a}\pp{b}}^{ab}-
\frac{109}{2520}  R  R_{ab}^{\pp{a}\pp{b}}  R_{\pp{a}\pp{b}}^{ab}+
\frac{73}{1890}  R_{ab}^{\pp{a}\pp{b}}  R_{c\pp{a}}^{\pp{c}a}  R_{\pp{b}\pp{c}}^{bc}+
\frac{1}{210}  R  R_{abcd}^{\pp{a}\pp{b}\pp{c}\pp{d}}  R_{\pp{a}\pp{b}\pp{c}\pp{d}}^{abcd}\nonumber \\ 
 &+
&\frac{1}{105}  R_{ab;cd}^{\pp{a}\pp{b}\pp{;\pp{c}}\pp{d}}  R_{\pp{a}\pp{c}\pp{b}\pp{d}}^{acbd}+
\frac{19}{630}  R_{ab}^{\pp{a}\pp{b}}  R_{cd}^{\pp{c}\pp{d}}  R_{\pp{a}\pp{c}\pp{b}\pp{d}}^{acbd}-
\frac{1}{189}  R_{abcd}^{\pp{a}\pp{b}\pp{c}\pp{d}}  R_{ef\pp{a}\pp{b}}^{\pp{e}\pp{f}ab}  
R_{\pp{c}\pp{d}\pp{e}\pp{f}}^{cdef},
                                                          \label{a3b}
\end{eqnarray}

\begin{eqnarray}
a_{3}^{(1)} &=& -
 \frac{1}{72} R^3-
\frac{1}{30}  R_{;a}^{\pp{;\pp{a}}}  R_{\pp{;\pp{a}}}^{;a}-
\frac{11}{180}  R  R_{;a\pp{a}}^{\pp{;\pp{a}}a} -
\frac{1}{180}  R  R_{abcd}^{\pp{a}\pp{b}\pp{c}\pp{d}}  R_{\pp{a}\pp{b}\pp{c}\pp{d}}^{abcd}
  \nonumber \\ 
 &-
&\frac{1}{60}  R_{;a\pp{a}b\pp{b}}^{\pp{;\pp{a}}a\pp{b}b}-
\frac{1}{90}  R_{;ab}^{\pp{;\pp{a}}\pp{b}}  R_{\pp{a}\pp{b}}^{ab}+
\frac{1}{180}  R  R_{ab}^{\pp{a}\pp{b}}  R_{\pp{a}\pp{b}}^{ab},
                                                        \label{a3c}
\end{eqnarray}

\begin{equation}
a_{3}^{(2)} = \frac{1}{12}R^3 +
\frac{1}{12}  R_{;a}^{\pp{;\pp{a}}}  R_{\pp{;\pp{a}}}^{;a}+
\frac{1}{6}  R  R_{;a\pp{a}}^{\pp{;\pp{a}}a}
                                                         \label{a3d}
\end{equation}
and

\begin{equation}
a_{3}^{(3)} = -
 \frac{1}{6} R^3.
                                                        \label{a3e}
\end{equation}
Before we proceed to the coefficient $[a_{4}]$ let us discuss briefly
this result. The package INVAR tries to expand each Riemann monomial 
in the basis of the independent Riemann invariants with no free indices ($r=0$).
This is achieved by defining polynomial relations between  dependent monomials
and the basis. Additionally, a number of dimensionally dependent identities
have been implemented. All this is of great practical importance since it is common
that small changes in the computational strategy can yield great differences 
in the theoretically equivalent results and the explicit demonstration of their equality
is extremely difficult.
Having at one's disposal a basis and rules provided by INVAR one can easily 
establish equivalence  of the results by construction a unique set of numerical 
coefficients and our calculations of $[a_3]$ and $[a_{4}]$ have been
compared and checked this way. 
It should be noted that the number of terms in $[a_{3}]$ has been reduced
as compared to the result presented, e.g., in Ref.~\cite{Leonard}, but, of course,
they are identical up to relatively simple identities satisfied by the Riemann tensor.

The coefficient $[a_{4}]$ is, on the other hand, extremely complicated and
even after massive simplifications it consists of 113 terms of
dimensionality of $[1/length]^{8}$. The coefficient $[a_{4}]$ can be written
in the form
\begin{equation}
[a_{4}] = a_{4}^{(0)} + \xi a_{4}^{(1)} + \xi^{2} a_{4}^{(2)} + \xi^3 a_{4}^{(3)}
+ \xi^{4} a_{4}^{(4)},
                                                      \label{a4a}
\end{equation}
where
\allowdisplaybreaks{
\begin{eqnarray}
a_{4}^{(0)} &=& \frac{1}{3780}  R_{;a\pp{a}b\pp{b}c\pp{c}}^{\pp{;\pp{a}}a\pp{b}b\pp{c}c} +
 \frac{5743}{1814400}  R^4+
\frac{229}{30240}  R  R_{;a}^{\pp{;\pp{a}}}  R_{\pp{;\pp{a}}}^{;a}-
\frac{241}{15120}  R  R_{ab;c}^{\pp{a}\pp{b}\pp{;\pp{c}}}  R_{\pp{a}\pp{b}\pp{;\pp{c}}}^{ab;c}-
\frac{1}{840}  R  R_{ab;c}^{\pp{a}\pp{b}\pp{;\pp{c}}}  R_{\pp{a}\pp{c}\pp{;\pp{b}}}^{ac;b}\nonumber \\ 
 &+         
&\frac{1}{672}  R  R_{abcd;e}^{\pp{a}\pp{b}\pp{c}\pp{d}\pp{;\pp{e}}}  R_{\pp{a}\pp{b}\pp{c}\pp{d}\pp{;\pp{e}}}^{abcd;e}+
\frac{1}{1800}  R_{;a\pp{a}}^{\pp{;\pp{a}}a}  R_{;b\pp{b}}^{\pp{;\pp{b}}b}+
\frac{13}{25200}  R_{;ab}^{\pp{;\pp{a}}\pp{b}}  R_{\pp{;\pp{a}}\pp{b}}^{;ab} -
\frac{1}{405}  R  R_{abcd}^{\pp{a}\pp{b}\pp{c}\pp{d}}  R_{ef\pp{a}\pp{b}}^{\pp{e}\pp{f}ab}  R_{\pp{c}\pp{d}\pp{e}\pp{f}}^{cdef}  \nonumber \\ 
 &+
&\frac{13}{37800}  R_{;ab}^{\pp{;\pp{a}}\pp{b}}  R_{\pp{a}\pp{b};c\pp{c}}^{ab\pp{;\pp{c}}c}-
\frac{1}{8400}  R_{ab;c\pp{c}}^{\pp{a}\pp{b}\pp{;\pp{c}}c}  R_{\pp{a}\pp{b};d\pp{d}}^{ab\pp{;\pp{d}}d}+
\frac{13}{12600}  R_{ab;cd}^{\pp{a}\pp{b}\pp{;\pp{c}}\pp{d}}  R_{\pp{a}\pp{b}\pp{;\pp{c}}\pp{d}}^{ab;cd}-
\frac{1}{450}  R_{ab;cd}^{\pp{a}\pp{b}\pp{;\pp{c}}\pp{d}}  R_{\pp{a}\pp{c}\pp{;\pp{b}}\pp{d}}^{ac;bd}\nonumber \\ 
 &+
&\frac{13}{12600}  R_{ab;cd}^{\pp{a}\pp{b}\pp{;\pp{c}}\pp{d}}  R_{\pp{c}\pp{d}\pp{;\pp{a}}\pp{b}}^{cd;ab}+
\frac{1}{3150}  R_{abcd;ef}^{\pp{a}\pp{b}\pp{c}\pp{d}\pp{;\pp{e}}\pp{f}}  R_{\pp{a}\pp{b}\pp{c}\pp{d}\pp{;\pp{e}}\pp{f}}^{abcd;ef}+
\frac{11}{7560}  R_{;a}^{\pp{;\pp{a}}}  R_{;b\pp{a}\pp{b}}^{\pp{;\pp{b}}ab}+
\frac{1}{1890}  R_{ab;c}^{\pp{a}\pp{b}\pp{;\pp{c}}}  R_{\pp{;\pp{a}}\pp{b}\pp{c}}^{;abc}\nonumber \\ 
 &-
&\frac{1}{7560}  R_{ab;c}^{\pp{a}\pp{b}\pp{;\pp{c}}}  R_{\pp{a}\pp{b}\pp{;\pp{c}}d\pp{d}}^{ab;c\pp{d}d}-
\frac{1}{3780}  R_{ab;c}^{\pp{a}\pp{b}\pp{;\pp{c}}}  R_{\pp{a}\pp{c}\pp{;\pp{b}}d\pp{d}}^{ac;b\pp{d}d}+
\frac{1}{315}  R_{abcd;e}^{\pp{a}\pp{b}\pp{c}\pp{d}\pp{;\pp{e}}}  R_{\pp{a}\pp{c}\pp{;\pp{b}}\pp{d}\pp{e}}^{ac;bde} +
\frac{13}{9450}  R_{abcd}^{\pp{a}\pp{b}\pp{c}\pp{d}}  R_{efg\pp{b}}^{\pp{e}\pp{f}\pp{g}b} 
 R_{h\pp{g}\pp{c}\pp{d}}^{\pp{h}gcd}  R_{\pp{a}\pp{h}\pp{e}\pp{f}}^{ahef}  \nonumber \\ 
 &-
&\frac{13}{15120}  R_{;a}^{\pp{;\pp{a}}}  R_{;b}^{\pp{;\pp{b}}}  R_{\pp{a}\pp{b}}^{ab}+
\frac{1}{4200}  R  R_{;ab}^{\pp{;\pp{a}}\pp{b}}  R_{\pp{a}\pp{b}}^{ab}-
\frac{5}{504}  R  R_{ab;c\pp{c}}^{\pp{a}\pp{b}\pp{;\pp{c}}c}  R_{\pp{a}\pp{b}}^{ab}
+
\frac{29}{75600}  R_{abcd}^{\pp{a}\pp{b}\pp{c}\pp{d}}  R_{efgh}^{\pp{e}\pp{f}\pp{g}\pp{h}}  
R_{\pp{a}\pp{b}\pp{c}\pp{d}}^{abcd}  R_{\pp{e}\pp{f}\pp{g}\pp{h}}^{efgh}
\nonumber \\ 
 &+
&\frac{1}{1890}  R_{;abc\pp{c}}^{\pp{;\pp{a}}\pp{b}\pp{c}c}  R_{\pp{a}\pp{b}}^{ab}-
\frac{1}{7560}  R_{ab;c\pp{c}d\pp{d}}^{\pp{a}\pp{b}\pp{;\pp{c}}c\pp{d}d}  R_{\pp{a}\pp{b}}^{ab}-
\frac{7253}{302400}  R^2  R_{ab}^{\pp{a}\pp{b}}  R_{\pp{a}\pp{b}}^{ab}-
\frac{271}{7560}  R_{;a}^{\pp{;\pp{a}}}  R_{bc\pp{;\pp{a}}}^{\pp{b}\pp{c};a}  R_{\pp{b}\pp{c}}^{bc} +
\frac{4}{1575}  R^2  R_{;a\pp{a}}^{\pp{;\pp{a}}a}  \nonumber \\ 
 &+
&\frac{1}{630}  R_{;a}^{\pp{;\pp{a}}}  R_{b\pp{a};c}^{\pp{b}a\pp{;\pp{c}}}  R_{\pp{b}\pp{c}}^{bc}-
\frac{17}{3150}  R_{;a\pp{a}}^{\pp{;\pp{a}}a}  R_{bc}^{\pp{b}\pp{c}}  R_{\pp{b}\pp{c}}^{bc}+
\frac{22}{675}  R  R_{ab}^{\pp{a}\pp{b}}  R_{c\pp{a}}^{\pp{c}a}  R_{\pp{b}\pp{c}}^{bc}+
\frac{89}{1890}  R_{ab;c}^{\pp{a}\pp{b}\pp{;\pp{c}}}  R_{d\pp{a}\pp{;\pp{c}}}^{\pp{d}a;c}  R_{\pp{b}\pp{d}}^{bd}\nonumber \\ 
 &+
&\frac{1}{378}  R_{ab;c}^{\pp{a}\pp{b}\pp{;\pp{c}}}  R_{d\pp{c}\pp{;\pp{a}}}^{\pp{d}c;a}  R_{\pp{b}\pp{d}}^{bd}+
\frac{83}{6300}  R_{ab;c\pp{c}}^{\pp{a}\pp{b}\pp{;\pp{c}}c}  R_{d\pp{a}}^{\pp{d}a}  R_{\pp{b}\pp{d}}^{bd}-
\frac{29}{4725}  R_{ab;cd}^{\pp{a}\pp{b}\pp{;\pp{c}}\pp{d}}  R_{\pp{a}\pp{c}}^{ac}  R_{\pp{b}\pp{d}}^{bd}-
\frac{341}{18900}  R_{ab}^{\pp{a}\pp{b}}  R_{cd}^{\pp{c}\pp{d}}  R_{\pp{a}\pp{c}}^{ac}  R_{\pp{b}\pp{d}}^{bd}\nonumber \\ 
 &-
&\frac{4}{945}  R_{ab;c}^{\pp{a}\pp{b}\pp{;\pp{c}}}  R_{d\pp{a}\pp{;\pp{b}}}^{\pp{d}a;b}  R_{\pp{c}\pp{d}}^{cd}+
\frac{1}{1260}  R_{ab;c}^{\pp{a}\pp{b}\pp{;\pp{c}}}  R_{\pp{a}\pp{b};d}^{ab\pp{;\pp{d}}}  R_{\pp{c}\pp{d}}^{cd}+
\frac{109}{18900}  R_{ab;cd}^{\pp{a}\pp{b}\pp{;\pp{c}}\pp{d}}  R_{\pp{a}\pp{b}}^{ab}  R_{\pp{c}\pp{d}}^{cd}+
\frac{703}{151200}  R_{ab}^{\pp{a}\pp{b}}  R_{cd}^{\pp{c}\pp{d}}  R_{\pp{a}\pp{b}}^{ab}  R_{\pp{c}\pp{d}}^{cd}\nonumber \\ 
 &+
&\frac{71}{3780}  R_{ab;c}^{\pp{a}\pp{b}\pp{;\pp{c}}}  R_{d\pp{a}e\pp{b}\pp{;\pp{c}}}^{\pp{d}a\pp{e}b;c}  R_{\pp{d}\pp{e}}^{de}+
\frac{1}{126}  R_{ab;c}^{\pp{a}\pp{b}\pp{;\pp{c}}}  R_{d\pp{a}e\pp{c}\pp{;\pp{b}}}^{\pp{d}a\pp{e}c;b}  R_{\pp{d}\pp{e}}^{de}-
\frac{1}{140}  R_{abcd;e}^{\pp{a}\pp{b}\pp{c}\pp{d}\pp{;\pp{e}}}  R_{f\pp{a}\pp{c}\pp{e}\pp{;\pp{b}}}^{\pp{f}ace;b}  R_{\pp{d}\pp{f}}^{df}-
\frac{1}{945}  R_{;a}^{\pp{;\pp{a}}}  R_{bc;d}^{\pp{b}\pp{c}\pp{;\pp{d}}}  R_{\pp{a}\pp{b}\pp{c}\pp{d}}^{abcd}\nonumber \\ 
 &+
&\frac{67}{33600}  R^2  R_{abcd}^{\pp{a}\pp{b}\pp{c}\pp{d}}  R_{\pp{a}\pp{b}\pp{c}\pp{d}}^{abcd}+
\frac{29}{6300}  R  R_{ab;cd}^{\pp{a}\pp{b}\pp{;\pp{c}}\pp{d}}  R_{\pp{a}\pp{c}\pp{b}\pp{d}}^{acbd}+
\frac{1}{945}  R_{ab;cde\pp{e}}^{\pp{a}\pp{b}\pp{;\pp{c}}\pp{d}\pp{e}e}  R_{\pp{a}\pp{c}\pp{b}\pp{d}}^{acbd}+
\frac{17}{18900}  R_{;ab}^{\pp{;\pp{a}}\pp{b}}  R_{cd}^{\pp{c}\pp{d}}  R_{\pp{a}\pp{c}\pp{b}\pp{d}}^{acbd}\nonumber \\ 
 &+
&\frac{83}{10800}  R  R_{ab}^{\pp{a}\pp{b}}  R_{cd}^{\pp{c}\pp{d}}  R_{\pp{a}\pp{c}\pp{b}\pp{d}}^{acbd}+
\frac{29}{1890}  R_{ab;c}^{\pp{a}\pp{b}\pp{;\pp{c}}}  R_{de\pp{;\pp{c}}}^{\pp{d}\pp{e};c}  R_{\pp{a}\pp{d}\pp{b}\pp{e}}^{adbe}+
\frac{32}{4725}  R_{ab;c\pp{c}}^{\pp{a}\pp{b}\pp{;\pp{c}}c}  R_{de}^{\pp{d}\pp{e}}  R_{\pp{a}\pp{d}\pp{b}\pp{e}}^{adbe}-
\frac{22}{4725}  R_{ab;cd}^{\pp{a}\pp{b}\pp{;\pp{c}}\pp{d}}  R_{e\pp{b}}^{\pp{e}b}  R_{\pp{a}\pp{d}\pp{c}\pp{e}}^{adce}\nonumber \\ 
 &+
&\frac{1}{4725}  R_{ab;cd}^{\pp{a}\pp{b}\pp{;\pp{c}}\pp{d}}  R_{e\pp{c}f\pp{d}}^{\pp{e}c\pp{f}d}  R_{\pp{a}\pp{e}\pp{b}\pp{f}}^{aebf}-
\frac{4}{945}  R_{ab;c}^{\pp{a}\pp{b}\pp{;\pp{c}}}  R_{de\pp{;\pp{b}}}^{\pp{d}\pp{e};b}  R_{\pp{a}\pp{e}\pp{c}\pp{d}}^{aecd}+
\frac{2}{1575}  R_{abcd;ef}^{\pp{a}\pp{b}\pp{c}\pp{d}\pp{;\pp{e}}\pp{f}}  R_{\pp{b}\pp{d}}^{bd}  R_{\pp{a}\pp{e}\pp{c}\pp{f}}^{aecf}  +
\frac{1}{1680}  R  R_{;a\pp{a}b\pp{b}}^{\pp{;\pp{a}}a\pp{b}b}  \nonumber \\ 
 &-
&\frac{1}{945}  R_{ab;c}^{\pp{a}\pp{b}\pp{;\pp{c}}}  R_{d\pp{a};e}^{\pp{d}a\pp{;\pp{e}}}  R_{\pp{b}\pp{c}\pp{d}\pp{e}}^{bcde}+
\frac{29}{7560}  R_{;a}^{\pp{;\pp{a}}}  R_{bcde\pp{;\pp{a}}}^{\pp{b}\pp{c}\pp{d}\pp{e};a}  R_{\pp{b}\pp{c}\pp{d}\pp{e}}^{bcde}+
\frac{1}{1400}  R_{;a\pp{a}}^{\pp{;\pp{a}}a}  R_{bcde}^{\pp{b}\pp{c}\pp{d}\pp{e}}  R_{\pp{b}\pp{c}\pp{d}\pp{e}}^{bcde}-
\frac{73}{9450}  R_{ab;cd}^{\pp{a}\pp{b}\pp{;\pp{c}}\pp{d}}  R_{ef\pp{a}\pp{d}}^{\pp{e}\pp{f}ad}  R_{\pp{b}\pp{c}\pp{e}\pp{f}}^{bcef}\nonumber \\ 
 &+
&\frac{1}{10800}  R_{ab}^{\pp{a}\pp{b}}  R_{cd}^{\pp{c}\pp{d}}  R_{ef\pp{a}\pp{c}}^{\pp{e}\pp{f}ac}  R_{\pp{b}\pp{d}\pp{e}\pp{f}}^{bdef}+
\frac{1}{315}  R_{ab;c}^{\pp{a}\pp{b}\pp{;\pp{c}}}  R_{def\pp{c}\pp{;\pp{a}}}^{\pp{d}\pp{e}\pp{f}c;a}  R_{\pp{b}\pp{f}\pp{d}\pp{e}}^{bfde}-
\frac{547}{302400}  R_{ab}^{\pp{a}\pp{b}}  R_{\pp{a}\pp{b}}^{ab}  R_{cdef}^{\pp{c}\pp{d}\pp{e}\pp{f}}  R_{\pp{c}\pp{d}\pp{e}\pp{f}}^{cdef}\nonumber \\ 
 &-
&\frac{1}{168}  R_{abcd;e}^{\pp{a}\pp{b}\pp{c}\pp{d}\pp{;\pp{e}}}  R_{fg\pp{a}\pp{b}\pp{;\pp{e}}}^{\pp{f}\pp{g}ab;e}  
R_{\pp{c}\pp{d}\pp{f}\pp{g}}^{cdfg}+
\frac{13}{18900}  R_{abcd}^{\pp{a}\pp{b}\pp{c}\pp{d}}  R_{efgh}^{\pp{e}\pp{f}\pp{g}\pp{h}}  
R_{\pp{a}\pp{b}\pp{e}\pp{f}}^{abef}  R_{\pp{c}\pp{d}\pp{g}\pp{h}}^{cdgh}+
\frac{103}{12600}  R_{ab}^{\pp{a}\pp{b}}  R_{cd}^{\pp{c}\pp{d}}  R_{e\pp{a}f\pp{b}}^{\pp{e}a\pp{f}b}  R_{\pp{c}\pp{f}\pp{d}\pp{e}}^{cfde},
\end{eqnarray}
                            \label{a4b}
}
\begin{eqnarray}
a_{4}^{(1)} &=& -
\frac{11}{1680}  R^4-
\frac{1}{112}  R  R_{;a}^{\pp{;\pp{a}}}  R_{\pp{;\pp{a}}}^{;a}+
\frac{1}{2520}  R  R_{ab;c}^{\pp{a}\pp{b}\pp{;\pp{c}}}  R_{\pp{a}\pp{b}\pp{;\pp{c}}}^{ab;c} +
\frac{1}{189}  R  R_{abcd}^{\pp{a}\pp{b}\pp{c}\pp{d}}  R_{ef\pp{a}\pp{b}}^{\pp{e}\pp{f}ab}  R_{\pp{c}\pp{d}\pp{e}\pp{f}}^{cdef}
  \nonumber \\ 
 &+
&\frac{1}{1260}  R  R_{ab;c}^{\pp{a}\pp{b}\pp{;\pp{c}}}  R_{\pp{a}\pp{c}\pp{;\pp{b}}}^{ac;b}-
\frac{1}{560}  R  R_{abcd;e}^{\pp{a}\pp{b}\pp{c}\pp{d}\pp{;\pp{e}}}  R_{\pp{a}\pp{b}\pp{c}\pp{d}\pp{;\pp{e}}}^{abcd;e}-
\frac{41}{5040}  R^2  R_{;a\pp{a}}^{\pp{;\pp{a}}a}-
\frac{1}{180}  R_{;a\pp{a}}^{\pp{;\pp{a}}a}  R_{;b\pp{b}}^{\pp{;\pp{b}}b}\nonumber \\ 
 &-
&\frac{1}{210}  R_{;ab}^{\pp{;\pp{a}}\pp{b}}  R_{\pp{;\pp{a}}\pp{b}}^{;ab}-
\frac{1}{630}  R_{;ab}^{\pp{;\pp{a}}\pp{b}}  R_{\pp{a}\pp{b};c\pp{c}}^{ab\pp{;\pp{c}}c}-
\frac{1}{72}  R_{;a}^{\pp{;\pp{a}}}  R_{;b\pp{a}\pp{b}}^{\pp{;\pp{b}}ab}-
\frac{1}{420}  R_{ab;c}^{\pp{a}\pp{b}\pp{;\pp{c}}}  R_{\pp{;\pp{a}}\pp{b}\pp{c}}^{;abc} -
\frac{1}{840}  R_{;a\pp{a}}^{\pp{;\pp{a}}a}  R_{bcde}^{\pp{b}\pp{c}\pp{d}\pp{e}}  R_{\pp{b}\pp{c}\pp{d}\pp{e}}^{bcde}  \nonumber \\ 
 &-
&\frac{2}{315}  R  R_{;a\pp{a}b\pp{b}}^{\pp{;\pp{a}}a\pp{b}b}-
\frac{1}{840}  R_{;a\pp{a}b\pp{b}c\pp{c}}^{\pp{;\pp{a}}a\pp{b}b\pp{c}c}+
\frac{17}{2520}  R_{;a}^{\pp{;\pp{a}}}  R_{;b}^{\pp{;\pp{b}}}  R_{\pp{a}\pp{b}}^{ab}-
\frac{1}{315}  R  R_{;ab}^{\pp{;\pp{a}}\pp{b}}  R_{\pp{a}\pp{b}}^{ab} -
\frac{1}{420}  R_{;a}^{\pp{;\pp{a}}}  
R_{bcde\pp{;\pp{a}}}^{\pp{b}\pp{c}\pp{d}\pp{e};a}  R_{\pp{b}\pp{c}\pp{d}\pp{e}}^{bcde}
  \nonumber \\ 
 &+
&\frac{1}{630}  R  R_{ab;c\pp{c}}^{\pp{a}\pp{b}\pp{;\pp{c}}c}  R_{\pp{a}\pp{b}}^{ab}-
\frac{1}{420}  R_{;abc\pp{c}}^{\pp{;\pp{a}}\pp{b}\pp{c}c}  R_{\pp{a}\pp{b}}^{ab}+
\frac{109}{2520}  R^2  R_{ab}^{\pp{a}\pp{b}}  R_{\pp{a}\pp{b}}^{ab}+
\frac{1}{1260}  R_{;a}^{\pp{;\pp{a}}}  R_{bc\pp{;\pp{a}}}^{\pp{b}\pp{c};a}  R_{\pp{b}\pp{c}}^{bc}\nonumber \\ 
 &+
&\frac{1}{630}  R_{;a}^{\pp{;\pp{a}}}  R_{b\pp{a};c}^{\pp{b}a\pp{;\pp{c}}}  R_{\pp{b}\pp{c}}^{bc}+
\frac{1}{504}  R_{;a\pp{a}}^{\pp{;\pp{a}}a}  R_{bc}^{\pp{b}\pp{c}}  R_{\pp{b}\pp{c}}^{bc}-
\frac{73}{1890}  R  R_{ab}^{\pp{a}\pp{b}}  R_{c\pp{a}}^{\pp{c}a}  R_{\pp{b}\pp{c}}^{bc}+
\frac{1}{630}  R_{;a}^{\pp{;\pp{a}}}  R_{bc;d}^{\pp{b}\pp{c}\pp{;\pp{d}}}  R_{\pp{a}\pp{b}\pp{c}\pp{d}}^{abcd}\nonumber \\ 
 &-
&\frac{1}{210}  R^2  R_{abcd}^{\pp{a}\pp{b}\pp{c}\pp{d}}  R_{\pp{a}\pp{b}\pp{c}\pp{d}}^{abcd}-
\frac{1}{105}  R  R_{ab;cd}^{\pp{a}\pp{b}\pp{;\pp{c}}\pp{d}}  R_{\pp{a}\pp{c}\pp{b}\pp{d}}^{acbd}-
\frac{1}{210}  R_{;ab}^{\pp{;\pp{a}}\pp{b}}  R_{cd}^{\pp{c}\pp{d}}  R_{\pp{a}\pp{c}\pp{b}\pp{d}}^{acbd}-
\frac{19}{630}  R  R_{ab}^{\pp{a}\pp{b}}  R_{cd}^{\pp{c}\pp{d}}  R_{\pp{a}\pp{c}\pp{b}\pp{d}}^{acbd},
                                                \label{a4c}
\end{eqnarray}
\begin{eqnarray}
a_{4}^{(2)} &=& \frac{1}{144}R^4 +
\frac{17}{360}  R  R_{;a}^{\pp{;\pp{a}}}  R_{\pp{;\pp{a}}}^{;a}+
\frac{2}{45}  R^2  R_{;a\pp{a}}^{\pp{;\pp{a}}a}+
\frac{1}{72}  R_{;a\pp{a}}^{\pp{;\pp{a}}a}  R_{;b\pp{b}}^{\pp{;\pp{b}}b}\nonumber \\ 
 &+
&\frac{1}{90}  R_{;ab}^{\pp{;\pp{a}}\pp{b}}  R_{\pp{;\pp{a}}\pp{b}}^{;ab}+
\frac{1}{30}  R_{;a}^{\pp{;\pp{a}}}  R_{;b\pp{a}\pp{b}}^{\pp{;\pp{b}}ab}+
\frac{1}{60}  R  R_{;a\pp{a}b\pp{b}}^{\pp{;\pp{a}}a\pp{b}b}-
\frac{1}{60}  R_{;a}^{\pp{;\pp{a}}}  R_{;b}^{\pp{;\pp{b}}}  R_{\pp{a}\pp{b}}^{ab}\nonumber \\ 
 &+
&\frac{1}{90}  R  R_{;ab}^{\pp{;\pp{a}}\pp{b}}  R_{\pp{a}\pp{b}}^{ab}-
\frac{1}{360}  R^2  R_{ab}^{\pp{a}\pp{b}}  R_{\pp{a}\pp{b}}^{ab}+
\frac{1}{360}  R^2  R_{abcd}^{\pp{a}\pp{b}\pp{c}\pp{d}}  R_{\pp{a}\pp{b}\pp{c}\pp{d}}^{abcd},
                                                \label{a4d}
\end{eqnarray}

\begin{eqnarray}
a_{4}^{(3)} &=& -
 \frac{1}{36}R^4-
\frac{1}{12}  R  R_{;a}^{\pp{;\pp{a}}}  R_{\pp{;\pp{a}}}^{;a}-
\frac{1}{12}  R^2  R_{;a\pp{a}}^{\pp{;\pp{a}}a}
                                                \label{a4e}
\end{eqnarray}
and

\begin{equation}
a_{4}^{(4)} = \frac{1}{24}R^4.
                                               \label{a4f}
\end{equation}
Although the total divergences can be discarded when working with the
effective action we have retained all the terms both in $[a_{3}]$ and
$[a_{4}],$ simply because they are interesting in their own right and can
be used in further calculations of, for example, the field fluctuation,
$\langle \phi^{2}\rangle.$

\section{Field fluctuation}

Before proceeding to the calculation of the renormalized stress-energy
tensor in the Reissner-Nordstr\"om geometry let us construct $\langle
\phi^{2}\rangle$ as it shares some of the general features of the full
stress-energy tensor while simultaneously being  calculationally less
involved. For example, if the next-to-leading term in the expansion leads
to substantial improvement of the result  it is likely that the quality of
the analytic approximation of the stress-energy tensor would also improve.
Similarly, if the vacuum polarization diverges on the event horizon it is
quite probable that the stress-energy tensor is also divergent there.

The vacuum polarization has been studied by a number of authors both in the
massive and massless cases. (See, e.g.~\cite{Novikov,Winstanley} and 
the references cited therein).
Generally, within the Schwinger-DeWitt framework the field fluctuation can
be expressed in terms of the coincidence limit of the coefficients $[a_{n}]$
and is given by \begin{equation}
\langle \phi^{2}\rangle_{k} = \frac{1}{16\pi^{2}}\sum_{n=2}^{k} \frac{(n-2)!}{m^{2(n-1)}}[a_{n}],
                         \label{fluctSq}
\end{equation}
where $k-1$ is the number of terms retained in the expansion. The leading
term of the expansion is to be constructed from the coefficient $[a_{2}],$
which, for the scalar field satisfying Eq.~(\ref{wave}) is given by
\begin{equation}
[a_{2}] = \frac{1}{180} R_{a b c d} R^{a b c d}-\frac{1}{180} R_{a b} R^{a b} +
\frac{1}{6}\left(\frac{1}{5}-\xi\right) R_{; a}^{\phantom{a} a}
+ \frac{1}{2}\left(\frac{1}{6} -\xi\right)^{2} R^{2}.
\end{equation} 
Having at our disposal compact expressions describing the first three
coefficients of the expansion (\ref{fluctSq}) we can  analyze the
influence of the higher order terms on the final result. Routine
calculations carried out in the Reissner-Nordstr\"om geometry give
\begin{equation}
[a_{2}] =\frac{1}{45r^{6}}\left(12 M^{2}-24\frac{MQ^{2}}{r} + 13\frac{Q^4}{r^{2}} \right),
\end{equation}
\begin{equation}
[a_{3}] =\frac{1}{r^{8}}\left(-\frac {194}{63}\,\frac {{M}^{3}}{r}+\frac {9}{7}\,M^{2}+
\frac {1111}{105}\,\frac {{M}^{2}{Q}^{2}}{{r}^{2}}-\frac {132}{35}
\,\frac {M{Q}^{2}}{r}-\frac {3644}{315}\,\frac {M{Q}^{4}}{{r}^{3
}}+\frac {908}{315}\,\frac {{Q}^{4}}{{r}^{2}}+\frac {1156}{315}
\,\frac {{Q}^{6}}{{r}^{4}}
 \right)
\end{equation}
and
\begin{eqnarray}
[a_{4}]& =& 
\frac{1}{r^{10}}\left({\frac {16549}{315}}\,{\frac {{Q}^{8}}{{r}^{6}}}+{\frac {50132}{675}}
\,{\frac {{Q}^{6}}{{r}^{4}}}-{\frac {373652}{1575}}\,{\frac {{Q}^{6}M}
{{r}^{5}}}+{\frac {571027}{1575}}\,{\frac {{Q}^{4}{M}^{2}}{{r}^{4}}}+{
\frac {612}{25}}\,{\frac {{Q}^{4}}{{r}^{2}}}
-{\frac {318476}{1575}}\,{
\frac {M{Q}^{4}}{{r}^{3}}}\right.
\nonumber \\
&+&\left. {\frac {3757}{25}}\,{\frac {{M}^{2}{Q}^{2}}
{{r}^{2}}}-{\frac {113522}{525}}\,{\frac {{M}^{3}{Q}^{2}}{{r}^{3}}}-{
\frac {4352}{175}}\,{\frac {M{Q}^{2}}{r}}+{\frac {32}{5}}\,{M}^{2}+{
\frac {23264}{525}}\,{\frac {{M}^{4}}{{r}^{2}}}-{\frac {240}{7}}\,{
\frac {{M}^{3}}{r}} \right).
\end{eqnarray}
Now, in order to gain insight into the nature of the thus constructed
approximations let us compare   $\langle \phi^{2}\rangle_{2}$ (the leading
term),  $\langle \phi^{2}\rangle_{3}$ ($\langle \phi^{2}\rangle_{2}$  plus
the next-to-leading term) and  $\langle \phi^{2}\rangle_{4}$ 
($\langle\phi^{2}\rangle_{3}$ plus the next-to-next-to-leading term). 
In Figs. 1 and 2  $\langle \phi^{2}\rangle_{k}$ for $k=2,3,4$ at the event horizon is
plotted against the admissible values of $|Q|/M.$ It is seen that the
second-order term considerably modifies the main approximation. On the
other hand, however, $\langle \phi^{2}\rangle_{3}$ and  $\langle 
\phi^{2}\rangle_{4}$  differ only slightly and consequently the changes in the
field fluctuation caused by $[a_{4}]$ are relatively small. Similarly, for
a given $Q$  the modification caused by the next-to-next-to-leading term is
small everywhere outside the event horizon.
Therefore,  it is reasonable to retain only the first two terms of the
expansion (\ref{fluctSq}). In what follows we shall demonstrate that a
similar pattern holds for the stress-energy tensor and the first two terms
of the expansion (\ref{Weff}) provide a good approximation. There is,
however, a profound difference between the two objects: to evaluate the
renormalized stress-energy tensor on has to retain the terms constructed
from $[a_{3}]$ and $[a_{4}]$ whereas analogous calculations of the field
fluctuation  require $[a_{2}]$ and $[a_{3}].$

\begin{figure}
 \includegraphics[width=9cm]{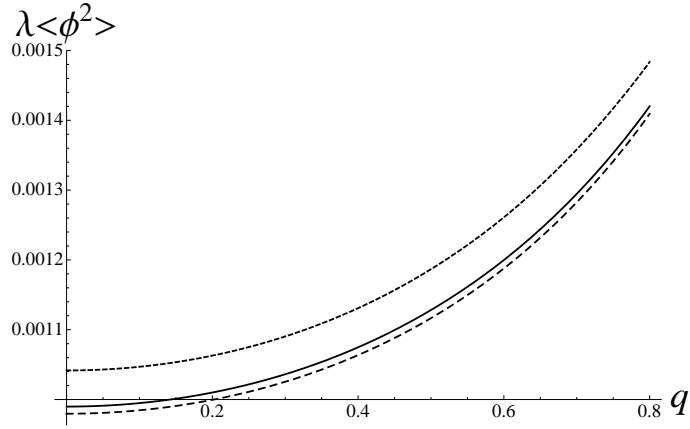}
 \caption{This graph shows the rescaled values of the vacuum polarization
$\langle \phi^{2}\rangle_{i}$ $[\lambda = 16 \pi^{2} M^{2}]$ at the event
horizon for the massive scalar field with $m M =2$  The solid line
corresponds to $\langle \phi^{2}\rangle_{4}$ the dashed line corresponds to
$\langle \phi^{2}\rangle_{3}$ and the dotted line corresponds to 
$\langle \phi^{2}\rangle_{2}.$
\label{fig1}}
 \end{figure}

\begin{figure}
 \includegraphics[width=9cm]{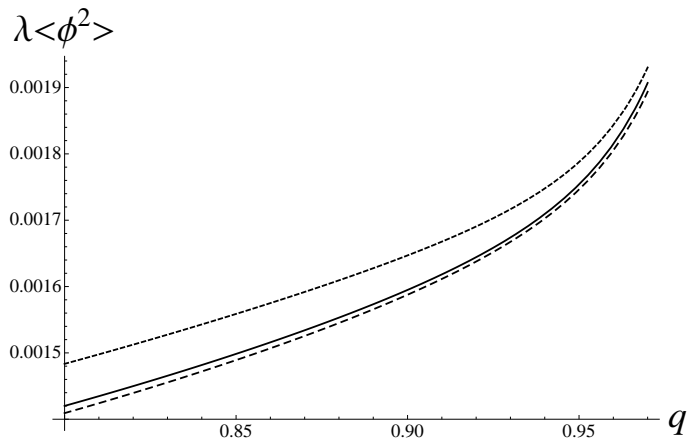}
 \caption{This graph shows the rescaled values of the vacuum polarization
$\langle \phi^{2}\rangle_{i}$ $[\lambda = 16 \pi^{2} M^{2}]$ at the event
horizon for the massive scalar field with $m M =2.$  The solid line
corresponds to $\langle \phi^{2}\rangle_{4}$ the dashed line corresponds to
$\langle \phi^{2}\rangle_{3}$ and the dotted line corresponds to 
$\langle \phi^{2}\rangle_{2}.$
\label{fig2}}
 \end{figure}

\section{The approximate stress-energy tensor}
\subsection{Effective action}

Having at one's disposal the approximate effective action, $W_{R},$ the
stress-energy tensor can be obtained form the standard formula
\begin{equation}
T^{ab} = \frac{2}{\sqrt{g}}\frac{\delta}{\delta g_{ab}} W_{R}.
                                           \label{funcDeriv}
\end{equation}
The total action that leads to the semiclassical Einstein field equations 
can be written in the form
\begin{equation}
S_{total} =\frac{1}{16\pi} \int R g^{1/2} d^{4}x + S_{m} + W_{R},
\end{equation}
where $S_{m}$ is the action of the classical sources and
\begin{equation}  
W_{R} = \frac{1}{32 \pi^{2} m^{2}} \int [a_{3}] g^{1/2} d^{4}x 
+\frac{1}{32 \pi^{2} m^{4}}\int [a_{4}] g^{1/2} d^{4}x.
\end{equation}
Now,  in order to calculate the functional derivatives of the action with
respect to the metric tensor the maximally simplified coefficients 
$[a_{3}(x,x')]$ and $[a_{4}(x,x')]$  have  once again been converted  into the
FORM language. This allows us to make use of our powerful FORM codes for
functional derivatives of the effective action. The final (mildly
simplified) result for the renormalized stress-energy tensor of the massive
scalar field with an arbitrary curvature coupling consists of a few thousand rather
complicated terms and is stored in the FORM format. It will not be
displayed here because it is too bulky. 
All our calculations in various geometries are
always carried out using this very tensor and the results are compared (if
necessary)  with the analogous results constructed using different (possibly
simpler) computational strategies. 

For the spherically symmetric line element expressed in the Schwarzschild
gauge 
\begin{equation} 
ds^{2} = f^{(0)}(r) dt^{2} + h^{(0)}(r) dr^{2} +
r^{2} \left(d\theta^{2} + \sin^{2} \theta d\phi^{2} \right). 
\end{equation}
one can save a lot of work by using the reduced action functionals. Indeed,
since the coefficients $[a_{n}]$ are constructed from various curvature
invariants they depend solely on the functions $f^{(0)}(r)$ and 
$h^{(0)}(r),$ their radial derivatives and the radial coordinate. Therefore, one
can easily perform simple integrations and reduce the problem to variations
with respect to the functions $f^{(0)}(r)$ and $h^{(0)}(r).$

The reduced action functional of the quantum part of the total action is
\begin{equation}
W_{R}^{reduced} =\frac{1}{32\pi^{2} m^{2}}\int dr [a_{3}] (f^{(0)} h^{(0)})^{1/2} r^{2} + 
\frac{1}{32\pi^{2} m^{4}}\int dr [a_{4}] (f^{(0)} h^{(0)})^{1/2} r^{2}.
                                               \label{lineelement}
\end{equation}
Although still tedious, this method requires substantially smaller number
of operations than the general one. This procedure yields $T^{tt}$ and
$T^{rr}$ components of the renormalized stress-energy tensor; the third
algebraically independent component, $T_{\theta}^{\theta} = T_{\phi}
^{\phi},$ can be obtained from the covariant conservation equation.

The quantum part of the total Lagrangian can schematically be written as
\begin{equation}
{\cal L} = {\cal L}_{1} + {\cal L}_{2},
                                          \label{Lagr_a}
\end{equation}
where
\begin{equation}
{\cal L}_{1} = {\cal L}_{1}\left(f^{(0)}(r),...,f^{(6)}(r),h^{(0)}(r),...,h^{(5)}(r),r\right)
                                          \label{Lagr_b}
\end{equation}
and
\begin{equation}
{\cal L}_{2} = {\cal L}_{2}\left(f^{(0)}(r),...,f^{(8)}(r),h^{(0)}(r),...,h^{(7)}(r),r\right).
                                          \label{Lagr_c}
\end{equation}
 $f^{(k)}$  and $h^{(k)}$ denote a $k-$th derivative of $f^{(0)}(r)$ and
$h^{(0)}(r),$ respectively. Note that the coefficients $(32\pi^{2} m^{2i})
^{-1}$ have been absorbed into the definition of ${\cal L}_{i}.$

The full form of ${\cal L}_{1}$ and ${\cal L}_{2}$ calculated for the line
element (\ref{lineelement}), when  expanded, consists of 531 and  2157
terms, respectively, and will not be presented here for obvious reasons. In
practice, there is no need to retain all the terms in Eqs. (\ref{a3a}) and
(\ref{a4a}). For example, for the $R =0$ class of metrics which is our main
interest here all the terms in Eqs. (\ref{a3d}), (\ref{a3e}) and
(\ref{a4d}-\ref{a4f})  do not contribute to the final result and the number
of terms in $a_{3}^{(0)},$ $a_{3}^{(1)}$ as well as $a_{4}^{(0)}$ and
$a_{4}^{(1)}$ is substantially reduced. Further simplifications can be 
obtained by neglecting the total divergences.

\subsection{Approximate stress-energy tensor in Reissner-Nordstr\"om geometry}
                                                 
Now the stress-energy tensor can be obtained from the Euler-Lagrange equations
\begin{equation}
T_{t}^{(i)t} =2 \left(\frac{f^{(0)}}{h^{(0)}}\right)^{1/2}
\left[ \frac{\partial}{\partial f^{(0)}}{\cal L}_{i} 
+  \sum_{k=1}^{n(i)}\left(-1\right)^{k+1} 
\frac{d^{k}}{dr^{k}} \left(\frac{\partial}{\partial f^{(k)}} {\cal L}_{i}\right) \right]
\end{equation}
and
\begin{equation}
T_{r}^{(i)r} =2 \left(\frac{h^{(0)}}{f^{(0)}}\right)^{1/2}
\left[\frac{\partial}{\partial h^{(0)}}{\cal L}_{i} 
+ \sum_{k=1}^{s(i)}\left(-1\right)^{k+1} \frac{d^{k}}{dr^{k}} 
\left(\frac{\partial}{\partial h^{(k)}} {\cal L}_{i}\right) \right],
\end{equation}
where $n(i)$ and $s(i)$ can easily be inferred form Eqs. (\ref{Lagr_b}) and
(\ref{Lagr_c}). The angular components can be easily obtained from the
covariant conservation equation $\nabla_{a} T^{ab} = 0,$ which, for the
line element (\ref{lineelement}), reduces to
\begin{equation}
T^{(i)\theta}_{\theta} = T^{(i)\phi}_{\phi} = -\frac{r}{4 f^{(0)}}
\left(T^{(i)t}_{t} -T^{(i)r}_{r} \right)\frac{d }{dr} f^{(0)} +
\frac{r}{2}\frac{d }{dr}T^{(i)r}_{r}  + T^{(i)r}_{r},
\end{equation} 
where $i = 1,2.$

Before we start calculations in the Reissner-Nordstr\"om geometry let us
discuss some features of the approximation to the stress-energy tensor
which can be deduced from the coefficients $[a_{3}]$ and $[a_{4}].$ First
observe that the stress-energy tensor depends linearly on the coupling
parameter $\xi$ and can be written as a sum of tensors of the type
\begin{equation}
T^{(i)b}_{a} =\frac{1}{\pi^{2} m^{2i} r^{2(3+i)}}\left( C^{(i)b}_{a} 
+ \eta D^{(i)b}_{a}\right),
\end{equation}
where $\eta = \xi -1/6.$ Further, it should be noted that since $T_{a}^{b}$
is constructed solely from the Riemann tensor and its covariant derivatives
it is regular for regular metrics. This property is guaranteed by the
polynomial character  of the result and the factorization:
\begin{equation}
T^{(i)t}_{t} -T^{(i)r}_{r} = f^{(0)}(r) P^{(i)}(r),
                                              \label{linearT}
\end{equation}
where $P^{(i)}(r)$ are the regular functions.

Although the first-order approximation to the stress-energy tensor in  the
Reissner-Nordstr\"om geometry is known, we shall display it for reader's
convenience. Making use of our general formulas, after some algebra,
we  conclude that the approximation has the form (\ref{linearT}), 
where  
\begin{eqnarray}
C^{(1)t}_{t} =
{\frac {3}{112}}\,{Q}^{2}-{\frac {13}{315}}\,{\frac {{Q}^{6}}{{r}^{4}}
}-{\frac {19}{672}}\,{M}^{2}-{\frac {4}{105}}\,{\frac {M{Q}^{2}}{r}}+{
\frac {313}{5040}}\,{\frac {{M}^{3}}{r}}-{\frac {101}{15120}}\,{\frac 
{{Q}^{4}}{{r}^{2}}}-{\frac {769}{10080}}\,{\frac {{M}^{2}{Q}^{2}}{{r}^
{2}}}+{\frac {257}{2520}}\,{\frac {M{Q}^{4}}{{r}^{3}}},
\end{eqnarray}

\begin{eqnarray}
D^{(1)t}_{t} =
-{\frac {11}{10}}\,{\frac {{M}^{3}}{r}}-\frac{7}{5}\,{\frac {M{Q}^{2}}{r}}+{
\frac {217}{60}}\,{\frac {{M}^{2}{Q}^{2}}{{r}^{2}}}-{\frac {113}{30}}
\,{\frac {M{Q}^{4}}{{r}^{3}}}+{\frac {91}{90}}\,{\frac {{Q}^{4}}{{r}^{
2}}}+{\frac {91}{80}}\,{\frac {{Q}^{6}}{{r}^{4}}}+\frac{1}{2}\,{M}^{2},
\end{eqnarray}

\begin{eqnarray}
C^{(1)r}_{r} =
{\frac {1}{96}}\,{M}^{2}+{\frac {3}{560}}\,{Q}^{2}+{\frac {421}{15120}
}\,{\frac {{Q}^{4}}{{r}^{2}}}-{\frac {31}{630}}\,{\frac {M{Q}^{2}}{r}}
-{\frac {11}{720}}\,{\frac {{M}^{3}}{r}}+{\frac {37}{2520}}\,{\frac {{
Q}^{6}}{{r}^{4}}}+{\frac {709}{10080}}\,{\frac {{M}^{2}{Q}^{2}}{{r}^{2
}}}-{\frac {23}{360}}\,{\frac {M{Q}^{4}}{{r}^{3}}},
\end{eqnarray}

\begin{eqnarray}
D^{(1)r}_{r} =
{\frac {7}{15}}\,{\frac {M{Q}^{2}}{r}}-{\frac {49}{60}}\,{\frac {{M}^{
2}{Q}^{2}}{{r}^{2}}}-{\frac {13}{45}}\,{\frac {{Q}^{4}}{{r}^{2}}}-\frac{1}{5}
\,{M}^{2}+{\frac {7}{10}}\,{\frac {M{Q}^{4}}{{r}^{3}}}-{\frac {13}{80}
}\,{\frac {{Q}^{6}}{{r}^{4}}}+\frac{3}{10}\,{\frac {{M}^{3}}{r}},
\end{eqnarray}

\begin{eqnarray}
C^{(1)\theta}_{\theta}& =& C^{(1)\phi}_{\phi}=
{\frac {407}{2520}}\,{\frac {M{Q}^{2}}{r}}-
\frac{M^{2}}{32} +{\frac {367}{5040}}\,{\frac {{M}^{3}}{r}}+{\frac {863}{
2520}}\,{\frac {M{Q}^{4}}{{r}^{3}}}-{\frac {761}{7560}}\,{\frac {{Q}^{
4}}{{r}^{2}}}-{\frac {367}{1120}}\,{\frac {{M}^{2}{Q}^{2}}{{r}^{2}}}-{
\frac {73}{720}}\,{\frac {{Q}^{6}}{{r}^{4}}} -{\frac {9Q^{2}}{560}}
\end{eqnarray}
and

\begin{eqnarray}
D^{(1)\theta}_{\theta}& =& D^{(1)\phi}_{\phi}= 
{\frac {91}{20}}\,{\frac {{M}^{2}{Q}^{2}}{{r}^{2}}}-\frac{7}{5}\,{\frac {{M}^{
3}}{r}}-{\frac {49}{30}}\,{\frac {M{Q}^{2}}{r}}-{\frac {71}{15}}\,{
\frac {M{Q}^{4}}{{r}^{3}}}+{\frac {52}{45}}\,{\frac {{Q}^{4}}{{r}^{2}}
}+\frac{3}{5}\,{M}^{2}+{\frac {117}{80}}\,{\frac {{Q}^{6}}{{r}^{4}}}.
\end{eqnarray}
This tensor is identical to that constructed in Refs.~\cite{AHS95,kocio1}.

Now, let us return to the tensor $T^{(2)b}_{a}.$ The calculations that lead
to this object are far more complicated than the analogous calculations of
the first-order term and require heavy use of the computer algebra. All
these efforts however will pay off and give us substantially better
approximation. In recent publications~\cite{kocio3,kocio2009b} it
has been argued that the minimal approximation constructed within the
framework of the Schwinger-DeWitt method should consist of the two first
terms of the expansion (\ref{Weff}). This observation was based on the
analyses carried out in the Schwarzschild and the Bertotti-Robinson
geometries. Specifically, it has been demonstrated  that the approximation
of the stress-energy tensor in the Schwarzschild spacetime constructed form
$[a_{3}]$ and $[a_{4}]$ is substantially better that the analogous
approximation calculated from the coefficient $[a_{3}]$ alone and this by
itself justifies the introduction of the second order term in that case. We
shall show that similar behavior occurs in the spacetime of the Reissner-
Nordstr\"om black hole. Moreover, the higher order terms may dramatically
change the type of the solutions of the semiclassical Einstein field
equations. An interesting example in this regard is given by the Bertotti-
Robinson geometry~\cite{robinson,bertotti}. Specifically, it can be shown that
although the Bertotti-Robinson geometry is a self-consistent solution of
the semiclassical Einstein field equations with the source term given
solely by the leading term of the renormalized stress-energy
tensor~\cite{Kofman1,Kofman2,kocio1,Olek} it does not remain so when the
next-to-leading term is taken into account. To guarantee that the Bertotti-
Robinson spacetime is the solution of the semiclassical equations one has
to introduce the (negative) cosmological constant. It should be noted that
addition of the electric charge to the system does not change this
behavior.
 
\begin{figure}
 \includegraphics[width=9cm]{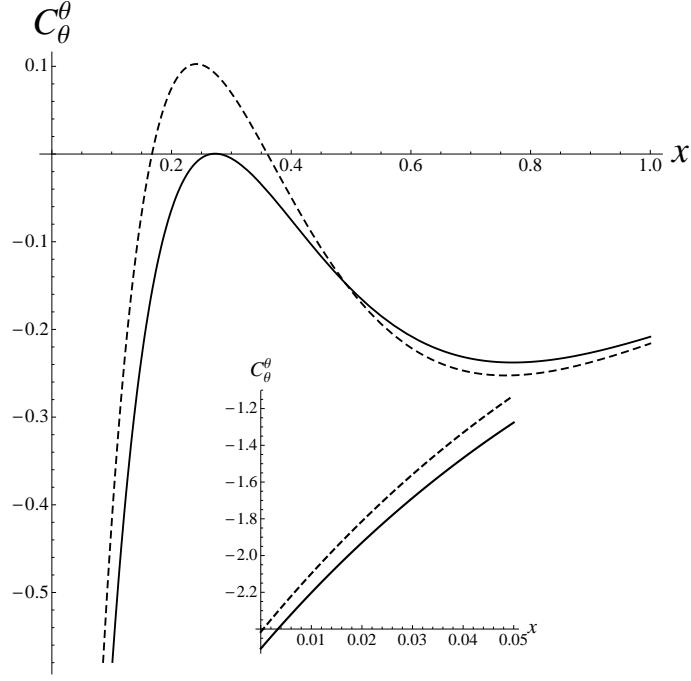}
 \caption{This graph shows the rescaled values of $C^{\theta}_{\theta}$
$[\lambda=90 (8M)^{4} \pi^{2}]$ as function of $x=(r-r_{+})/M$ for massive
scalar field with $m M =2$ and $|Q|/M=0.95.$ The solid line corresponds to
the improved approximation whereas the dashed line to the first order
Schwinger-DeWitt approximation. In the small panel the near horizon
behavior of $C^{\theta}_{\theta}$ is displayed
\label{fig3}}
 \end{figure}

Thee second-order term has, as expected, the form (\ref{linearT}),
where
\begin{eqnarray}
C^{(2)t}_{t} &=&-{\frac {11}{40}}\,{M}^{2}+\frac{1}{12}\,{Q}^{2}
-{\frac {255229}{151200}}\,{
\frac {{Q}^{8}}{{r}^{6}}}+{\frac {2833}{2100}}\,{\frac {{M}^{3}}{r}}+{
\frac {41063}{6300}}\,{\frac {M{Q}^{4}}{{r}^{3}}}+{\frac {157}{350}}\,
{\frac {M{Q}^{2}}{r}}-{\frac {409}{525}}\,{\frac {{Q}^{4}}{{r}^{2}}}\nonumber \\
&&-{
\frac {13583}{8400}}\,{\frac {{M}^{4}}{{r}^{2}}}+{\frac {95839}{12600}
}\,{\frac {M{Q}^{6}}{{r}^{5}}}-{\frac {190013}{75600}}\,{\frac {{Q}^{6
}}{{r}^{4}}}-{\frac {6131}{1400}}\,{\frac {{M}^{2}{Q}^{2}}{{r}^{2}}}-{
\frac {287009}{25200}}\,{\frac {{M}^{2}{Q}^{4}}{{r}^{4}}}+{\frac {
83611}{12600}}\,{\frac {{M}^{3}{Q}^{2}}{{r}^{3}}},
\end{eqnarray}

\begin{eqnarray}
D^{(2)t}_{t} &=&{\frac {1649}{56}}\,{\frac {{M}^{4}}{{r}^{2}}}+{\frac {56361}{560}}\,{
\frac {{M}^{2}{Q}^{2}}{{r}^{2}}}-{\frac {2785}{21}}\,{\frac {M{Q}^{4}}
{{r}^{3}}}+\frac{9}{2}\,{M}^{2}-{\frac {47}{2}}\,{\frac {{M}^{3}}{r}}+{\frac {
390577}{1680}}\,{\frac {{M}^{2}{Q}^{4}}{{r}^{4}}}
\nonumber \\
&&-{\frac {2251}{15}}\,
{\frac {M{Q}^{6}}{{r}^{5}}}-{\frac {594}{35}}\,{\frac {M{Q}^{2}}{r}}-{
\frac {71087}{504}}\,{\frac {{M}^{3}{Q}^{2}}{{r}^{3}}}+{\frac {227}{14
}}\,{\frac {{Q}^{4}}{{r}^{2}}}+{\frac {20207}{420}}\,{\frac {{Q}^{6}}{
{r}^{4}}}+{\frac {41327}{1260}}\,{\frac {{Q}^{8}}{{r}^{6}}},
\end{eqnarray}

\begin{eqnarray}
C^{(2)r}_{r} &=&
{\frac {3}{40}}\,{M}^{2}+{\frac {1}{84}}\,{Q}^{2}+{\frac {34463}{
151200}}\,{\frac {{Q}^{8}}{{r}^{6}}}-{\frac {97}{300}}\,{\frac {{M}^{3
}}{r}}-{\frac {7897}{4725}}\,{\frac {M{Q}^{4}}{{r}^{3}}}-{\frac {239}{
700}}\,{\frac {M{Q}^{2}}{r}}+{\frac {247}{945}}\,{\frac {{Q}^{4}}{{r}^
{2}}}
\nonumber \\
&&
+{\frac {2753}{8400}}\,{\frac {{M}^{4}}{{r}^{2}}}-{\frac {151}{
120}}\,{\frac {M{Q}^{6}}{{r}^{5}}}+{\frac {4321}{8400}}\,{\frac {{Q}^{
6}}{{r}^{4}}}+{\frac {1531}{1050}}\,{\frac {{M}^{2}{Q}^{2}}{{r}^{2}}}+
{\frac {11581}{5040}}\,{\frac {{M}^{2}{Q}^{4}}{{r}^{4}}}-{\frac {19907
}{12600}}\,{\frac {{M}^{3}{Q}^{2}}{{r}^{3}}},
\end{eqnarray}

\begin{eqnarray}
D^{(2)r}_{r} &=&
-{\frac {291}{56}}\,{\frac {{M}^{4}}{{r}^{2}}}-{\frac {34127}{1680}}\,
{\frac {{M}^{2}{Q}^{2}}{{r}^{2}}}+{\frac {14911}{630}}\,{\frac {M{Q}^{
4}}{{r}^{3}}}-{\frac {9}{7}}\,{M}^{2}+{\frac {151}{28}}\,{\frac {{M}^{
3}}{r}}-{\frac {3487}{112}}\,{\frac {{M}^{2}{Q}^{4}}{{r}^{4}}}
\nonumber \\
&&
+{\frac 
{254}{15}}\,{\frac {M{Q}^{6}}{{r}^{5}}}
+{\frac {297}{70}}\,{\frac {M{Q
}^{2}}{r}}+{\frac {1567}{72}}\,{\frac {{M}^{3}{Q}^{2}}{{r}^{3}}}-{
\frac {227}{63}}\,{\frac {{Q}^{4}}{{r}^{2}}}-{\frac {9433}{1260}}\,{
\frac {{Q}^{6}}{{r}^{4}}}-{\frac {3757}{1260}}\,{\frac {{Q}^{8}}{{r}^{
6}}},
\end{eqnarray}

\begin{eqnarray}
C^{(2)\theta}_{\theta}& =& C^{(2)\phi}_{\phi}= 
-\frac{3}{10}\,{M}^{2}-\frac{1}{21}\,{Q}^{2}-{\frac {386087}{151200}}\,{\frac {{Q}^{8}
}{{r}^{6}}}+{\frac {163}{100}}\,{\frac {{M}^{3}}{r}}+{\frac {55159}{
5400}}\,{\frac {M{Q}^{4}}{{r}^{3}}}+{\frac {2101}{1400}}\,{\frac {M{Q}
^{2}}{r}}
-{\frac {961}{756}}\,{\frac {{Q}^{4}}{{r}^{2}}}
\nonumber \\
&&
-{\frac {17849
}{8400}}\,{\frac {{M}^{4}}{{r}^{2}}}+{\frac {73417}{6300}}\,{\frac {M{
Q}^{6}}{{r}^{5}}}-{\frac {137681}{37800}}\,{\frac {{Q}^{6}}{{r}^{4}}}-
{\frac {16657}{2100}}\,{\frac {{M}^{2}{Q}^{2}}{{r}^{2}}}-{\frac {18259
}{1008}}\,{\frac {{M}^{2}{Q}^{4}}{{r}^{4}}}+{\frac {138431}{12600}}\,{
\frac {{M}^{3}{Q}^{2}}{{r}^{3}}}
\end{eqnarray}
and

\begin{eqnarray}
D^{(2)\theta}_{\theta}& =& D^{(2)\phi}_{\phi}= 
\frac {485}{14}\,{\frac {{M}^{4}}{{r}^{2}}}+{\frac {38663}{336}}\,{
\frac {{M}^{2}{Q}^{2}}{{r}^{2}}}-{\frac {189871}{1260}}\,{\frac {M{Q}^
{4}}{{r}^{3}}}+{\frac {36}{7}}\,{M}^{2}-{\frac {1521}{56}}\,{\frac {{M
}^{3}}{r}}
+{\frac {151701}{560}}\,{\frac {{M}^{2}{Q}^{4}}{{r}^{4}}}
\nonumber \\
&&
-{
\frac {18446}{105}}\,{\frac {M{Q}^{6}}{{r}^{5}}}-{\frac {2673}{140}}\,
{\frac {M{Q}^{2}}{r}}-{\frac {41513}{252}}\,{\frac {{M}^{3}{Q}^{2}}{{r
}^{3}}}+{\frac {1135}{63}}\,{\frac {{Q}^{4}}{{r}^{2}}}+{\frac {9869}{
180}}\,{\frac {{Q}^{6}}{{r}^{4}}}+{\frac {48841}{1260}}\,{\frac {{Q}^{
8}}{{r}^{6}}}.
\end{eqnarray}
The angular components of the renormalized stress-energy tensor of the
massive scalar field in the Reissner-Nordstr\"om geometry has been
calculated numerically for $m M =2$ and $|Q|/M=0.95.$  (Similar
calculations have also been carried out for $|Q|=0.$) However, in view of
further applications, it is preferable to have at one's disposal simple and
accurate general analytic formulas describing the functional dependence of
the stress-energy tensor on the metric. Such general formulas can easily be
applied to the concrete line element provided some general requirements
concerning the geometry and the mass of the quantized  field are satisfied.
Consequently, it is of interest to compare the approximation constructed
from $[a_{3}]$ and $[a_{4}]$ with the results of the numerical calculations
of the conformal and nonconformal contribution to the total stress-energy
tensor
\begin{equation}
T_{\theta}^{\theta} =C_{\theta}^{\theta} + \left(\xi -\frac{1}{6} 
\right) D_{\theta}^{\theta}.
\end{equation}
as presented in ref.~\cite{AHS95}

 \begin{figure}
 \includegraphics[width=9cm]{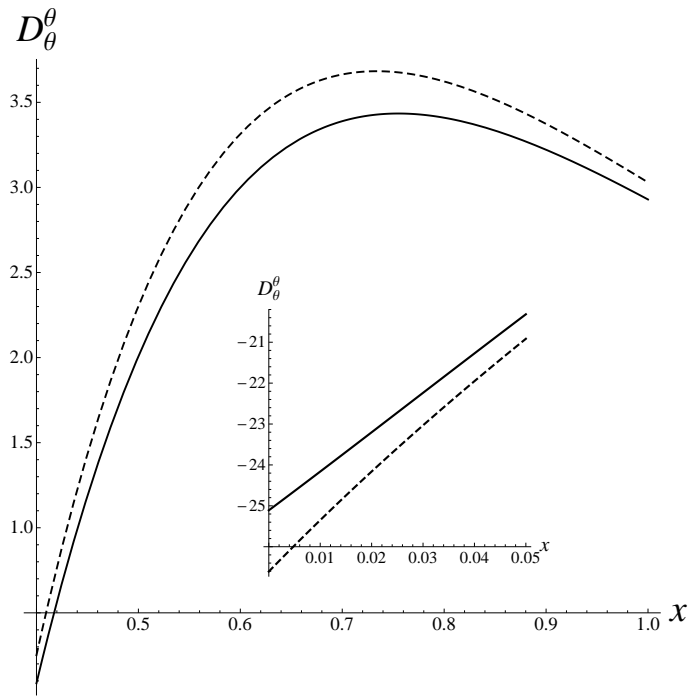}
 \caption{This graph shows the rescaled values of $D^{\theta}_{\theta}$
$[\lambda=90 (8M)^{4} \pi^{2}]$ as function of $x=(r-r_{+})/M$ for massive
scalar field with $m M =2$ and $|Q|/M=0.95.$ The solid line corresponds to
the improved approximation whereas the dashed line to the first order
Schwinger-DeWitt approximation. In the small panel the near horizon
behavior of $D^{\theta}_{\theta}$ is displayed.
\label{fig4}}
 \end{figure}

A comparison of Figs. 3 and 4 of the present paper  with the Figs. 10 and
11 of Ref.~\cite{AHS95} clearly shows that although the first order
approximation correctly reproduces qualitative behavior of $C_{\theta}
^{\theta}$ and $D_{\theta}^{\theta}$, the inclusion of the next-to-leading
term substantially improves the approximation of the stress-energy tensor
even in the closest vicinity of the event horizon. One expects that this
approximation is even better for $m M>2.$ A lesson that follows from this
demonstration is that the next-to-leading term plays, or at least may play,
an important role in the calculations and it can be ignored only after
careful examination. Similar behavior of the next-to-leading term of 
the stress-energy tensor  in the Schwarzschild spacetime has been reported in Ref.~\cite{kocio3}.
One expects therefore  that  this pattern holds for all
values satisfying $0\leq|Q|/M \leq 0.95.$ There is also  good reason 
to believe that it is true for all admissible values of $q.$

At the event horizon of the extreme Reissner-Nordstr\"om black hole 
the stress-energy tensor can be written in the remarkably simple form
\begin{equation}
T_{a}^{(1)b} =\frac{1}{\pi^{2} m^{2} M^{6}} \left(\frac{1}{3780} -\frac{\eta}{720}  \right)diag [1,1,-1,-1],
\end{equation}
\begin{equation}
T_{a}^{(2)b}= -\frac{11}{151200\pi^{2} m^{4} M^{8}} diag[1,1,1,1].
\end{equation}
As the geometry in the vicinity of the degenerate Reissner-Nordstr\"om
solution is precisely that of the Bertotti-Robinson one can easily
calculate the approximate stress-energy tensor. Indeed, due to homogeneity
($R_{abcd;e}=0$), vanishing of the Ricci scalar and the  Weyl tensor ($C_{abcd} =0$),
the stress-energy tensor can be expressed solely in terms of the Ricci
tensor.  Making use of the general formulas the first order approximation
to the stress-energy tensor  can be written in the form
\begin{eqnarray}
32\pi^{2} m^{2} T^{(1) i j} &=&
-\frac{5 - 14 \xi}{1260} R_{a b}R^{a b} R^{i j},
\label{tep1}
\end{eqnarray}
whereas when the above conditions are satisfied the next-lo-leading term reads
\begin{eqnarray}
32\pi^{2} m^{2}  T^{(2)i j}&=&
-\frac{11}{75600} R_{a b} R^{a b} R_{c d}R^{c d} g^{i j}.
\label{tep2}
\end{eqnarray}
Simple calculation shows that (\ref{tep1}) and (\ref{tep2}) in the
Bertotti-Robinson geometry are precisely equivalent to the stress-energy
tensor at the degenerate horizon of the extreme Reissner-Nordstr\"om black
hole and this may be regarded as the additional useful check of the
calculations.

\section{Final remarks}

In this work our goal was to construct the  approximate field fluctuation
and  renormalized stress-energy tensor of the quantized massive field in
the spacetime of the Reissner-Nordstr\"om black hole and to investigate how
the higher-order terms of the expansions (\ref{Weff}) and (\ref{fluctSq}) 
affect the final results. 
The general formulas describing the both quantities are extremely
complex, but, fortunately, there are massive simplifications 
when applied to the static and spherically symmetric geometries.
A comparison with the numeric calculations
reported in a classic paper by Anderson, Hiscock and Samuel~\cite{AHS95} shows that the
next-to-leading term substantially improves the approximation. It has been
found that in both cases the minimal approximations are to be constructed
from the first two terms of (\ref{fluctSq})  and (\ref{Weff}) for the field 
fluctuation and the stress-energy tensor, respectively.
Although we have constructed the general form of the  stress-energy tensor 
up to the next-to-leading terms by functional differentiation of the action 
functional with respect to the metric,  here we proposed a computationally simpler 
method in which  the reduced action functionals are varied
with respect to the functions $g_{tt}(r)$ and $g_{rr}(r).$
Both methods, when overlap,  give, of course, identical results.

We hope that our results will be of use in further calculations. We
indicate a few possible directions of investigations. First, it would be
interesting to analyze the back reaction of the quantized massive field
upon the geometry of the Reissner-Nordstr\"om black hole. 
Due to simplicity of the stress-energy tensor in the Reisssner-Nordstr\"om 
spacetime the quantum-corrected metric can easily be constructed performing 
two elementary quadratures. Since the quantum
part in the right  hand side of the semiclassical Einstein field equations
is calculated in a large mass limit it is purely geometric quantity and can be expressed
solely in terms of the Riemann tensor, its covariant derivatives and contractions.
This allows to treat the semiclassical theory as the higher derivative theory
and construct various characteristics encoded in the geometry of the quantum-corrected 
black hole such as location of the horizons in nondegenerate as well as degenerate 
case~\cite{Dobado,Wise}, equations of motion of the test particles~\cite{Wise,Gosia},
temperature and entropy~\cite{Wald1,Iyer,Visser1,Visser2,Gosia,Ja}
and energy-momentum complexes~\cite{Ja2}.
Further, construction of the approximate stress-energy tensor as well as the field
fluctuation in more complex backgrounds and the accompanying numerical calculations
would certainly strengthen our understanding of the problem. Especially
interesting in this regard is the problem of the lukewarm~\cite{Winstanley}
and ultraextremal~\cite{Oleg} black holes. Finally, an important and
interesting continuation of the calculations presented in this paper would be
construction of the of the next-to-leading term of the  spinor and vector
fields. We intend to return to this group of problems elsewhere.


\end{document}